Ferromagnetism at 300 K in spin-coated films of Co doped anatase and rutile-TiO$_2$

R. Suryanarayanan[1*], V.M. Naik[2], P. Kharel[1], P. Talgala[1] and R. Naik[1]

[1]Department of Physics and Astronomy, Wayne State University, 666 W Hancock, Detroit, MI 48201, USA

[2]Department of Natural Sciences, University of Michigan-Dearborn, Dearborn, MI, 48128, USA

Thin films of Ti$_{1-x}$Co$_x$O$_2$ (x=0 and 0.03) have been prepared on sapphire substrates by spin-on technique starting from metalorganic precursors. When heat treated in air at 550 and 700 C respectively, these films present pure anatase and rutile structures as shown both by X-ray diffraction and Raman spectroscopy. Optical absorption indicate a high degree of transparency in the visible region. Such films show a very small magnetic moment at 300 K. However, when the anatase and the rutile films are annealed in a vacuum of 1x10$^{-5}$ Torr at 500 $^o$C and 600 $^o$C respectively, the magnetic moment, at 300 K, is strongly enhanced reaching 0.36 $\mu_B$/Co for the anatase sample and 0.68 $\mu_B$/Co for the rutile one. The ferromagnetic Curie temperature of these samples is above 350 K.

[*]Permanent address: LPCES, UMR8648, ICMMO, Universite Paris-Sud, 91405 Orsay, France. E mail: sury39@yahoo.com

Transparent ferromagnetic semiconductors with ferromagnetic Curie temperature, T$_c$>300 K, are expected to play an important role in the development of spintronics [1]. Earlier reports based on III-V semiconductors such as GaAs, doped with Mn, revealed that T$_c$ hardly exceeded 150 K rendering them unsuitable for viable room temperature applications [2]. Since then, the attention has turned towards the oxide semiconductors. Matsumoto et al., [3] was the first to report the occurrence of ferromagnetism with T$_c$>300 K in Co doped TiO$_2$ films obtained by pulsed laser deposition (PLD). Following this, several reports have appeared using different preparative techniques such as oxygen-plasma-assisted molecular-beam epitaxy (MBE) [4], sputtering [5], ion implantation [6], laser MBE [7] as well as PLD [8-10]. Though all these samples did show unambiguous signs of ferromagnetic behavior at 300 K or above, it was difficult to exclude the possibility of formation of Co clusters and thus attribute the observed magnetism entirely to Co substituted TiO$_2$. Several arguments, for and against the formation of clusters, were proposed by these authors. To this date, this continues to be a hot debate.

We have used an entirely different non-vacuum approach, known as spin-on technique, to prepare Co substituted $TiO_2$ films. This is a well-known technique that has been used to prepare $TiO_2$ films starting from metalorganic precursors [11]. The main advantages of this technique are low cost, simplicity, high purity of starting materials and the possibility of a homogeneous mixture of two cations in the liquid state. We report here on the successful preparation of 3% Co substituted $TiO_2$ films crystallizing either in anatase or rutile structure and present X-ray diffraction (XRD), optical absorption, Raman spectra and magnetization measurements. Though the as prepared films showed a ferromagnetic behavior at 300 K, the vacuum annealed films showed a strong enhancement in the magnetic moment of Co.

The metalorganic precursors used were Ti-ethylhexoxide and Co-neodecanoate. The solutions were thoroughly mixed in approximately 95 to 5 ratio (by volume) using an ultrasonic bath. A small amount of xylene was added to obtain optimal viscosity needed for spin-coating. A few drops of this mixed solution were placed on a sapphire substrate (c-axis oriented of 25 mm$^2$ area) which was spun at 5000 rpm for 15 sec. The sample was annealed in air at 550 $^{\circ}$C for one minute and the process was repeated 5 to 10 times to build up the layer thickness. The thickness of the sample used in the present study was 800 nm. Several such samples were prepared and were subjected to different heat treatments in air. The EDX showed that the concentration of Co was about 3%. Films were further vacuum annealed under $1 \times 10^{-5}$ Torr at 500 $^{\circ}$C or 600 $^{\circ}$C.

The as prepared films with and without Co heat treated at 550 $^{\circ}$C in air show a pure anatase phase (Fig. 1a) whereas the films annealed at 700 $^{\circ}$C in air show a rutile structure, as revealed by XRD (Fig. 1b). The films are polycrystalline with no preferred orientation. The different pure phases of the films were further confirmed by Raman spectra as shown in Fig. 2a,b, which clearly exhibit all the characteristic phonon modes, expected for the two forms of $TiO_2$. These are in full agreement with the spectra observed in single crystalline forms of rutile and anatase-$TiO_2$ both of which belong to $D_{4h}$ space group [12, 13]. The broad band ~235 cm$^{-1}$ in rutile-$TiO_2$ (Fig. 2b) has been assigned as a combination line rather than a fundamental one phonon process [13]. The 612 cm$^{-1}$ ($A_{1g}$) phonon region also has combination bands in the vicinity making the room temperature $A_{1g}$ mode broad and asymmetric. The mode at 144 cm$^{-1}$ ($B_{1g}$) is expected to be very weak in pure rutile phase. Figures 3a(i) and (ii) and Fig. 3b(i) and (ii) show the Raman spectra of $Ti_{0.97}Co_{0.03}O_2$ before and after vacuum annealing in the case of anatase and rutile films, respectively. The spectra of $Ti_{0.97}Co_{0.03}O_2$ in the anatase form are identical to that of pure anatse-$TiO_2$ indicating that the anatase structure is retained after doping with Co. This suggests that dopants occupy the substitutional sites in the host lattice. Further, no noticeable change in the spectra are seen due to the vacuum annealing. In the case of rutile-$Ti_{0.97}Co_{0.03}O_2$ films, the $B_{1g}$ mode has somewhat larger intensity compared to the pure rutile form. We note that, in the powder form of $TiO_2$, the presence of $B_{1g}$ mode in the rutile phase has been interpreted as an indication of presence of a small fraction of anatase phase (<5%) [14, 15]. We also note that the presence of any unreacted cobalt oxides is not indicated by our data.

The optical absorption spectra of anatase and rutile-$TiO_2$ films are shown in Fig.4(a). The observed intensity oscillations, in 450 – 800 nm region of the spectra are due to optical interference effects and the features observed < 400 nm are assigned as due to electronic bandgap transitions. Anatase- $TiO_2$ has an absorption edge ~ 350 nm, whereas the rutile-$TiO_2$ has an absorption edge ~ 375 nm. The optical absorption edges remain same after doping with Co (Fig.4b). The observed absorption edges in the present work are consistent with the reported values in the literature for the two forms of $TiO_2$. We did not notice any appreciable change in the optical spectra of these two samples after they were annealed in vacuum.

The magnetization (M) of the Co substituted samples before and after vacuum annealing was measured as a function of temperature and magnetic field (H) with the help of a Quantum Design superconducting quantum interference device (SQUID) magnetometer. M as a function of H (-5000 to 5000 Oe) at 300 K of the blank substrate was also measured in order to eliminate the substrate contribution. The data discussed here thus represent the magnetization of the films. Fig. 5 shows M as a function of H at 300 K of anatase sample before and after annealing. Several interesting features can be noted. First, at 300 K, the as prepared anatase form of Co substituted $TiO_2$ film shows a very small magnetic moment of 0.15 $\mu_B$/Co in a field of 2000 Oe. Second, the vacuum annealed sample at 500 $^oC$, show a strong enhancement in the ferromagnetic properties, the magnetic moment reaching a value of 0.36 $\mu_B$/Co. Also, the coercivity increases from a few Oe to 480 Oe as a result of this annealing. The remanent magentization ($M_r$) at 300 K is 0.22 $\mu_B$/Co. The M of the rutile sample is also very sensitive to the vacuum annealing treatment but the behavior is different from that of the anatase sample (Fig. 6). The value of M of the as prepared rutile sample is quite small (0.08 $\mu_B$/Co ) in a field of 2000 Oe but increases to 0.2 $\mu_B$/Co after annealing at 500 $^oC$. The same sample after annealing at 600 $^oC$, shows a remarkable increase in M to 0.68 $\mu_B$/Co, which is higher than that of the anatase sample but annealed at a lower temperature of 500 $^oC$. Note that the antase sample is not stable if annealed at temperatures higher than 500 $^oC$. Whereas the coercivity of the 500 $^oC$ annealed rutile sample remains almost unaltered at a value of 80 Oe, a remarkable increase to a value of 325 Oe is observed after it was annealed at 600 C. The value of $M_r$ is 0.25 $\mu_B$/Co. Further, M as a function of temperature, measured in a field of 2000 Oe, in the case of both the anatse and the rutile samples, after annealing at 500 and 600 C respectively is fairly constant from 5 K to 350K (Figs. 5 and 6 inset). Since the temperature range of our magnetization measurements is limited to 350 K, we expect our samples to remain ferromagnetic at least up to 400 K. The data of our samples are comparable to those prepared by PLD technique [10].

In what follows, we present a short discussion of our data in comparison with those published earlier and further comment briefly on a recent model that has been proposed by Coey et al., [16] and Venkatesan et al., [17] to account for the occurrence of ferromagnetism in transparent semiconductors. To start with, we note that the earlier reports suggested that the retention of the anatase or rutile phase depended on the nature of the substrate. For example, Hong et al., [10] observed the anatase phase when $LaAlO_3$ and $SrTiO_3$ substrates were used whereas the films deposited on the Si substrates showed the rutile phase. Others have obtained either the anatase phase [4] or the rutile phase [7]. On the contrary, it is interesting to note that we could obtain either the anatase phase or the rutile phase on the sapphire substrates just by selecting a proper heat treatment, as revealed unambiguously by Raman data. And further, the pure phases are retained after the heat treatment in vacuum. Next we address the question regarding the formation of magnetic clusters. Murakami et al.,[7] concluded from the EXAFS, XANES and XPS studies that their rutile films contained both Co metallic clusters and Co substituted into Ti site. Kim et al. [18] came to a similar conclusion in their anatase samples by studying X-ray magnetic circular dichroism. On the other hand, Hong et al., [10] by conducting magnetic force microscope studies did not observe any nanometer-sized clusters on the surface. Further, they point out that the observed value of the magnetic moment of 0.23 $\mu_B$/Co in their samples is small compared to the value of 1.7$\mu_B$/Co expected for Co metal. Our values of 0.36 $\mu_B$/Co for the anatase sample and 0.68 $\mu_B$/Co for the rutile sample are also relatively smaller comapred to that expected for Co metal. However, recent results show that one can obtain even much higher values of magnetic moment of Co in $SnO_2$ [ref.19], or ZnO [17] or $HfO_2$ [ref.20]. For example, Ogale et al., [19] have observed 7.5 $\mu_B$/Co in Co doped $SnO_2$ films and Ramachandra Rao et al., [20] have observed 6.8 $\mu_B$/Co in Co doped $HfO_2$ samples. Note that these authors do not attribute the higher value to the presence of clusters but to the unquenched orbital moment of Co. In the absence of an exchange mechanism which could account for the observed data at low doping levels, Coey et al., [16] have proposed a model where the exchange is mediated by carriers in a spin-split impurity band derived from extended donor orbitals. They point out that the oxygen vacancies play an important role. Though the details have not been worked out yet, this model is very appealing to us since annealing of our films in vacuum, could possibly create oxygen vacancies that could influence the magnetic properties in this diluted system. In the present case, it may also induce another valence state of Ti, assuming Co is in a valence state of 2. Further, the value of the magnetic moment of our anatase sample is different from that of the rutile sample when both of them were annealed at 500 $^oC$ in vacuum. This may arise from the different defect chemistry of these two phases resulting from the heat treatments. In fact, the rutile sample does show a remarkable increase in the value of the magnetic moment when it is further annealed at 600 $^oC$. Detailed studies on the effect of annealing on the magnetic properties and the microstructure of the TiO$_2$ doped with Co and other elements seem to be highly desirable in order to understand further this interesting aspect. Such studies are expected ~~out~~ to throw further light on the role played by oxygen defects on the magnetic properties.

In summary, by using a low cost spin-on technique, we have successfully prepared Co($\approx$3%) substituted TiO$_2$ films on sapphire substrates both in the anatase and in the rutile forms that show room temperature ferromagnetism. The magnetic properties are found to be enhanced as a result of annealing in vacuum. Upon annealing at 500 $^o$C, the magnetic moment at 300 K of the anatase sample is 0.36 $\mu_B$/Co whereas that of the rutile sample is 0.2 $\mu_B$/Co. However, the latter value of the rutile sample increases to 0.68 $\mu_B$/Co when annealed at 600 $^o$C. Our data, though cannot rule out the presence of Co clusters, point out the important role played by oxygen defects.

Figure Captions

Fig. 1   XRD $\theta-2\theta$ scans of (a) anatase and (b) rutile TiO$_2$ films

Fig. 2   Raman spectrum of (a) anatase and (b) rutile TiO$_2$ films. The band marked with an asterisk is a combination band.

Fig. 3   Raman spectra of Ti$_{0.97}$Co$_{0.03}$O$_2$ films in (a) anatase and (b) in rutile forms. Spectra labeled (*i*) samples without vacuum annealing, and (*ii*) samples after vacuum annealing at 500 $^{o}$C for 30 minutes. The asterisks represent the peaks originating from the sapphire substrates.

Fig. 4   Optical transmission spectra of (a) pure TiO$_2$ and (b) Ti$_{0.97}$Co$_{0.03}$O$_2$ after vacuum annealing. The spectra of Ti$_{0.97}$Co$_{0.03}$O$_2$ before and after vacuum annealing are identical.

Fig. 5   Magnetic Moment vs Magnetic field at 300 K for Anatase Ti$_{0.97}$Co$_{0.03}$O$_2$ film (symbols o and ● represent the data before and after vacuum annealing). The inset shows the Magnetic Moment vs Temperature data in a field of 2000 Oe.

Fig. 6   Magnetic Moment vs Magnetic field at 300 K for Rutile Ti$_{0.97}$Co$_{0.03}$O$_2$ film (symbols o and ● represent the data before and after vacuum annealing at 500 $^{o}$C. Symbol $\Delta$ represents the data after vacuum annealing at 600 $^{o}$C and the inset shows the corresponding magnetic moment vs temperature data in a field of 2000 Oe).

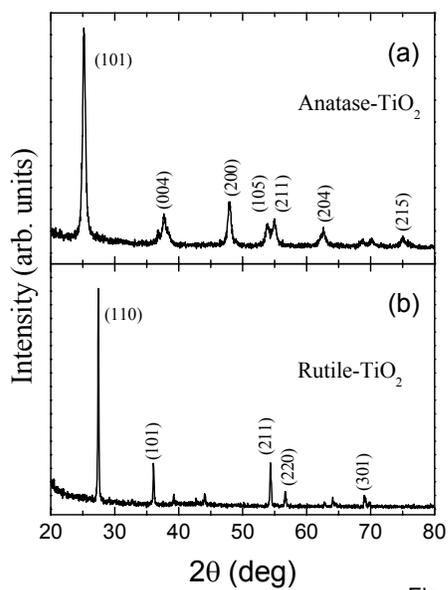

Fig. 1

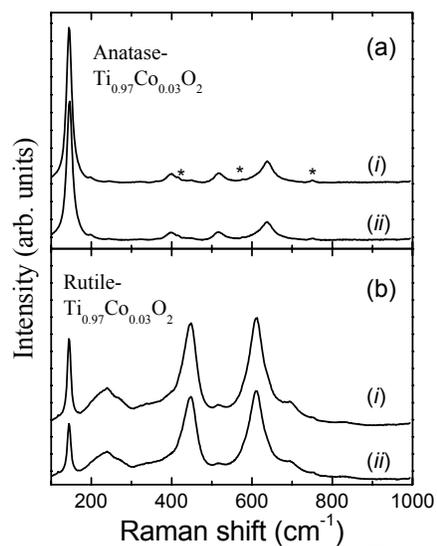

Fig. 3

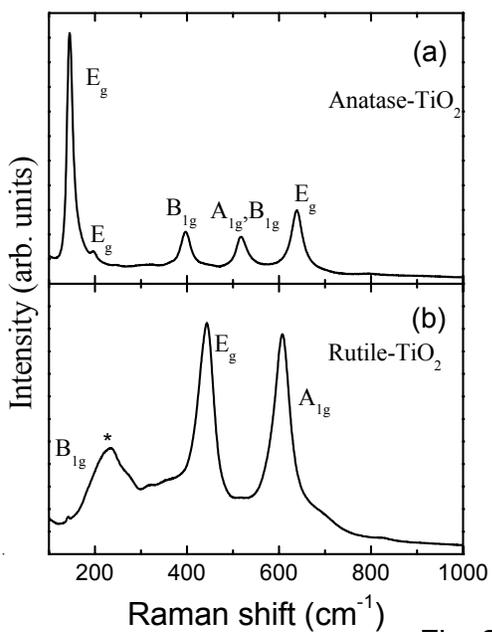

Fig. 2

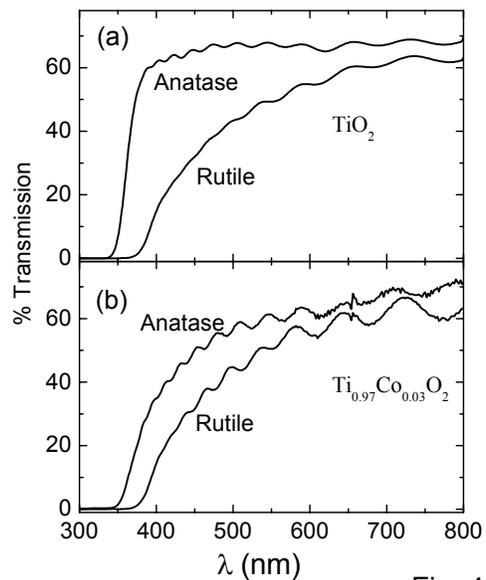

Fig. 4

Fig. 5

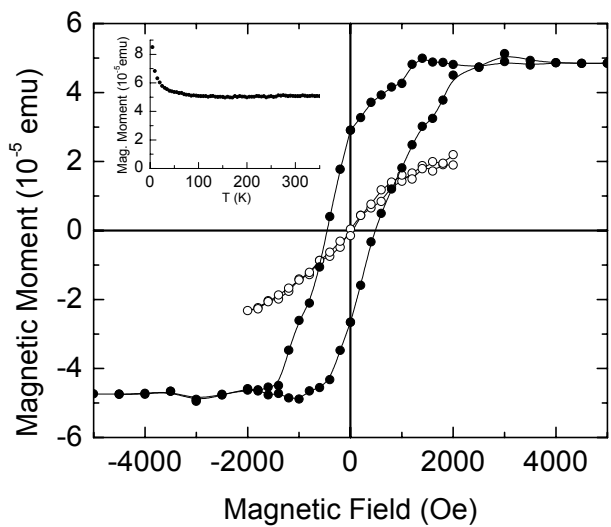

Fig. 6

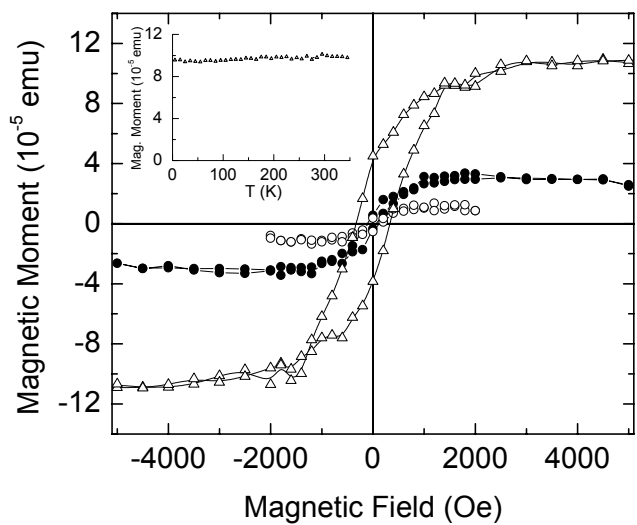